\documentclass[12pt,draftcls,journal,leterpaper,twosides,onecolumn]{IEEEtran}
\usepackage[english]{babel}
\usepackage{amsmath,amssymb,amscd,latexsym,dsfont}
\usepackage{float,color,graphicx,subfigure,balance}
\usepackage{multirow,multicol}
\usepackage{comment,cite}
\usepackage{enumerate}
\usepackage{stfloats}
\usepackage{psfrag}
\usepackage{cite}
\usepackage[colorlinks=true,linkcolor=blue]{hyperref} 

\IEEEoverridecommandlockouts

\title{Linear Correction of Mismatched L-values \\ in BICM receivers}
\author{%
\authorblockN{Leszek Szczecinski\\}
\authorblockA{ INRS-EMT, Montreal, Canada\\}
\emph{leszek@emt.inrs.ca}

\thanks{The work was supported by the government of Quebec, under grant \#PSR-SIIRI-435. Parts of this work have been submitted to IEEE International Conference on Communications 2012.}
}%

\newcommand{\tr}[1]{\textrm{#1}}
\newcommand{\mb}[1]{\mathbf{#1}}
\newcommand{\ov}[1]{\overline{#1}}

\newcommand{\mc}[1]{\mathcal{#1}}

\newcommand{\ms}[1]{\mathds{#1}}

\newcommand{\bc}{\boldsymbol{c}}

\newcommand{\set}[1]{\{#1\}}
\newcommand{\bset}[1]{\left\{#1\right\}}

\newcommand{\cd}{\cdot}
\newcommand{\ld}{\ldots}

\newtheorem{theorem}{Theorem}

\newtheorem{proposition}{Proposition}

\newtheorem{example}{Example}

\begin{document}

\maketitle

\begin{abstract}
In this work we analyze the problem of linear correction of the reliability metrics (L-values) in BICM receivers. We want to find the correction factors that minimize the probability of error of a maximum likelihood decoder that uses the corrected L-values. To this end, we use the efficient approximation of the pairwise error probability in the domain of the cumulant generating functions (CGF) of the L-values and conclude that the optimal correction factors are equal to the twice of the saddlepoint of the CGF. We provide a simple numerical example of transmission in the presence of interference where we demonstrate a notable improvement attainable with the proposed method. The proposed method is compared with the one  based on the maximization of generalized mutual information.
\end{abstract}


\section{Introduction}\label{Sec:Intro}

The logarithmic likelihood ratios (LLRs, or L-values) calculated at the receiver for the transmitted bits, are a convenient representation of  the likelihood of the observations and are often used in all of the processing operations in the receiver (such as ``soft'' detection, decoding, iterative processing, etc). In this work we consider the so-called \emph{mismatched} L-values, which only approximate the true L-values and to ``correct'' the mismatch, that may occur due to many independent reasons, we analyze the linear scaling of the mismatched L-values. Formulating the problem in the context of BICM receivers, we aim at the minimization of the probability of error of  the maximum likelihood (ML) decoder that uses the corrected L-values.

The  L-value $l_{n}$ of the bit $C_{n}$ (transmitted at time $n$) is a well known way of representing the reliability of the transmitted bit. It is related to the observation $y_{n}$ via
\begin{align}\label{LLR.intro}
   l_{n}=\log\frac{p_{Y_{n}|C_{n}}(y_{n}| 1)}{p_{Y_{n}|C_{n}}(y_{n}| 0)},
\end{align}
where $p_{Y_{n}|C_{n}}(y_{n}| b)$ is the probability density function (pdf) of the observation $Y_{n}$ conditioned on the sent bit $C_{n}=b$. 

The L-values are basic signals/messages exchanged between the processing units. The multiplications of probabilities required in many processing steps transform into addition of corresponding L-values; the numerical simplicity of the resulting operations is the reason behind the popularity of the L-values. For example,  in BICM receivers, the L-values are calculated by the front-end detector and then passed to the decoder \cite{Zehavi92}. In some cases, operations on the L-values are carried out before decoding as it happens when combining the signals obtained in independent transmissions of the same bit \cite{Wengerter02}. The L-values are also used in binary decoders that operate in an iterative fashion, e.g., turbo-decoders \cite{Benedetto97} or message passing algorithms used for decoding of LDPC codes \cite{Chen02b}. 

In some situations, however, the L-values are not appropriately calculated, or are \emph{mismatched}. Ignoring the mismatch when processing the L-values is, in general, suboptimal and to correct it, nonlinear operations on the L-values may be required. To make the correction simple, a linear operation (i.e., multiplication by a correction factor) is often considered. This idea was already studied in the context of BICM receivers \cite{Classon02}, turbo-decoding \cite{Papke96}\cite{Vogt00}, or LDPC decoding \cite{Chen02b}. However, the correction factor was most often found through a brute-force search, that is, among the results obtained for different correction factors the one ensuring the best performance is deemed optimal. While this is a pragmatic approach when searching for one or two correction factors, it cannot be applied when many correction factors have to be found (the search space becomes too large) and/or when the correction has to be done on-line (i.e., when it depends on many continuously varying parameters).

The works in \cite{Papke96}\cite{Dijk03}\cite{Lechner06b}\cite{Nguyen11}\cite{Yazdani11} aimed at finding the correction factor using the pdf of the L-value. The method of \cite{Papke96}, based on a Gaussian model of the L-value fails to capture properties of non-Gaussian pdfs while \cite{Dijk03} draws general conclusions about the suitability of linear correction but relies on simulation to find the correction factor. \cite{Lechner06b}\cite{Nguyen11}\cite{Yazdani11} rely on the minimization of a functional of the pdf which requires numerical integration as in most cases the analytical solutions are not available. The drawbacks of  \cite{Lechner06b} is that the pdf has to be known or estimated and the functional in the optimization problem is not related to any performance criterion. This disadvantages were recently removed in \cite{Nguyen11}\cite{Yazdani11}, where the correction factor was formally found via maximization of the so-called generalized mutual information (GMI) between the L-values and the corresponding bits. Then, even if the pdf is not known, the Monte-Carlo integration can be implemented. While this approach was (experimentally) shown to improve the performance of BICM receivers operating with the capacity-approaching codes, it does not explicitly address the problem of minimizing the error probability of the optimal maximum-likelihood (ML) decoder.

In this paper we explicitly aim at the minimization of the probability of error in ML decoders, which results in a novel correction principle and provides a new insight into correction of the L-values.  Our problem is formulated in the domain of the cumulative generating function (CGF) of the L-values. As their calculation is simpler than finding the pdf, in many cases we will be able to avoid explicit numerical integration. We find a simple correction principle which says that the correction factor should be equal to the twice of the so-called saddlepoint of the CGF, which is the real argument  minimizing the CGF. Finding the saddlepoint requires solving a simple non-linear equation which, in many cases, may be even found analytically.

The paper is organized as follows. The definitions and notation are presented in Sec.~\ref{Sec:Model} and the new correction principle we propose is explained in Sec.~\ref{Sec:Min.PEP}. A detailed illustration of our analysis in shown in Sec.~\ref{Sec:Num.Results} on an example of correction of the L-values in the BICM receivers operating in the presence of interference. 

\section{Model}\label{Sec:Model}

We consider a scenario where a codeword of  $N$ bits $\bc=[c_{1}, c_{2},\ld, c_{N}]$ is sent over a binary-input memoryless channel with known transition probability given by the pdf $p_{Y_{n}|C_{n}}(y_{n}|c_{n})$.

Upon reception of $y_{1}, \ld, y_{N}$, in order to minimize the probability of detection error, the  decoder decides in favour of the codeword that maximizes the likelihood of the observation, i.e., 
\begin{align}
\hat{\bc}
              &=\tr{argmax}_{\bc\in\mc{C}}\sum_{n=1}^{N}\log p_{Y_{n}|C_{n}}(y_{n}|c_{n}), \label{ML.dec}
\end{align}
where $\mc{C}$ is the code (i.e., set of all codewords).

Using Bayes' formula $p_{Y_{n}|C_{n}}(y_{n}|c_{n})= P_{C_{n}|Y_{n}}(c_{n}|y_{n})p_{Y_{n}}(y_{n})/P_{C_{n}}(c_{n})$ in \eqref{LLR.intro}, and knowing that $P_{C_{n}|Y_{n}}(c_{n}|y_{n})=1-P_{C_{n}|Y_{n}}(1-c_{n}|y_{n})$, we obtain a useful alternative expression of the aposteriori probabability $P_{C_{n}|Y_{n}}(c_{n}|y_{n})=e^{l_{n}\cd c_{n}}/(1+e^{l_{n}})$. It transforms \eqref{ML.dec} into the decoding based on the L-values
\begin{align}
\hat{\mb{c}}&=\tr{argmax}_{\mb{c}\in\mc{C}}\sum_{n=1}^{N} l_{n}c_{n} , \label{ML.dec.LLR}
\end{align}
where the terms independent of $\bc$ were removed from the maximization in \eqref{ML.dec.LLR}.

The L-values $l_{n}$ are modelled as random variables $L_{n}$ and if they are calculated exactly as defined in \eqref{LLR.intro} their pdf satisfies the so-called \emph{consistency condition} \cite[Sec.~III]{Tuchler04}
\begin{align}\label{cons.cond}
  \frac{p_{L_{n}|C_{n}}(l|1)}{p_{L_{n}|C_{n}}(l|0)}=\tr{e}^{l}.
\end{align}
The so-called \emph{symmetry condition} \cite[Sec.~III]{Tuchler04}
\begin{align}\label{symm.cond}
p_{L_{n}|C_{n}}(l|c)=p_{L_{n}|C_{n}}(-l| 1-c)
\end{align}
simplifies the analysis and, although it does not have to be always satisfied (it depends on the conditional pdf $p_{Y_{n}|C_{n}}(y_{n}|c_{n})$), it may be forced by a pseudo-random scrambling of the bits $c_{n}$ prior to modulation,  followed by the change of the sign of the L-values $l_{n}$ if the bit was negated \cite{Caire98}. 
Thus, from now on, we assume that this condition is always satisfied.

Rewriting \eqref{cons.cond} as $p_{L_{n}|C_{n}}(l|1)\tr{e}^{-l/2}=\tr{e}^{l/2}p_{L_{n}|C_{n}}(l|0)$ and using \eqref{symm.cond} yields what we call a \emph{consistency-symmetry} condition
 \begin{align}
  p_{L_{n}|C_{n}}(l|0)\tr{e}^{l/2}&=\tr{e}^{-l/2}p_{L_{n}|C_{n}}(-l|0).\label{cons.symm.cond}
\end{align}

\subsection{Mismatched decoding and correction of L-values}\label{Sec:Mismatch.Dec}
In practice, the calculation of some L-values via \eqref{LLR.intro} may be inexact because i)~the model $p_{Y_{n}|C_{n}}(y_{n}|c_{n})$ is not accurate, ii)~its parameters are not well estimated, or iii)~the likelihood is calculated using simplifications introduced to lower the computational effort. In general, these effects may be represented as if a ``mismatched'' likelihood $q(y_{n},c_{n})\neq p_{Y_{n}|C_{n}}(y_{n}|c_{n})$ was used  in \eqref{LLR.intro} yielding the ``mismatched'' L-values \cite{Martinez09}\cite{Nguyen11}
\begin{align}\label{LLR.mismatch}
\tilde{l}_{n}=\log\frac{q(y_{n},1)}{q(y_{n},0)}.
\end{align}

If the mismatch is ignored, that is, $\tilde{l}_{n}$ is falsely assumed to be identical with $l_{n}$, the receiver will operate in a suboptimal fashion because $\tilde{l}_{n}$ cannot be transformed into the likelihood $p_{Y_{n}|C_{n}}(y_{n}|c_{n})$. Nevertheless, if the conditional distribution $p_{\tilde{L}_{n}| C_{n}}( \tilde{l} | c)$ of $\tilde{L}$ (that models the mismatched metrics $\tilde{l}$) is known, we might calculate a post-processing or ``correction'' function \cite{Dijk03}\cite{Lechner06b}
\begin{align}\label{}
  f^\tr{c}( l )=\log\frac{p_{\tilde{L}_{n}| C_{n}}( l | 1)}{p_{\tilde{L}_{n}| C_{n}}( l | 0)},
\end{align}
and then, calculated the ``corrected'' L-value as $\tilde{l}^\tr{c}=f^\tr{c}\bigl(\tilde{l}_{n}\bigr)$.

In general, the effect of the mismatch cannot be eliminated, i.e., $\tilde{l}^\tr{c}_{n}\neq l_{n}$. However, using $\tilde{l}^\tr{c}_{n}$ instead of $\tilde{l}_{n}$ should improve the performance of the decoder,   because $\tilde{l}^\tr{c}_{n}$ \emph{does} represent the likelihood of the observation $\tilde{l}_{n}$ conditioned on the bit $c_{n}$. 
We also immediately conclude that if the L-value is matched, i.e., its pdf satisfies \eqref{cons.cond}, no correction is necessary as we obtain $f^\tr{c}(l)=l$, that is, $\tilde{l}^\tr{c}_{n}=\tilde{l}_{n}$.

\begin{example}
If we assume the Gaussian form of the pdf $p_{\tilde{L}_n|C_{n}}(l|0)=\Psi\left(l+\tilde{\mu},\tilde{\sigma}^{2}\right)=p_{\tilde{L}_n|C_{n}}(-l|1)$, where 
\begin{align}\label{Gauss.function}
\Psi( l; \sigma^{2})=\frac{1}{\sqrt{2\pi}\sigma}\exp\left(-\frac{l^{2}}{2\sigma^{2}}\right),
\end{align}
$\tilde{\mu}=-\ms{E}_{\tilde{L}_{n}|C_{n}=0}\set{\tilde{L}_{n}}$ is the negated mean of $\tilde{L}_{n}$ and $\tilde{\sigma}^{2}=\ms{E}_{\tilde{L}_{n}|C_{n}=0}\set{\tilde{L}_{n}^{2}}-\ms{E}^{2}_{\tilde{L}_{n}|C_{n}=0}\set{\tilde{L}_{n}}$ its variance,  the correction function
\begin{align}\label{alpha.Gauss}
f^\tr{c}(l)=\frac{2\tilde{\mu}}{\tilde{\sigma}^{2}}\cd l =\hat{\alpha}^\tr{Gauss}\cd l.
\end{align}
is linear and the correction factor is determined by the double of the ratio of the mean and the variance of the L-value. 
\end{example}

The Gaussian model from the above example was used in \cite{Papke96} to justify the correction based on $\hat{\alpha}^\tr{Gauss}$.

The resulting correction $\tilde{l}^\tr{c}=\alpha\cd \tilde{l}$ has an appealing simplicity and in many cases (treated mostly in the area of iterative decoding) $f^\tr{c}(l)$ was observed to be relatively well approximated by a linear function  \cite{Dijk03}\cite{Lechner06b}\cite{Alvarado09b}. Therefore, using $f(l)=\alpha l$ instead of $f^\tr{c}(l)$ (that is, in general, nonlinear) often provides the satisfactory correction effect \cite{Dijk03}\cite{Lechner06b}\cite{Alvarado09b} and, when compared to non-linear functions $f^\tr{c}(l)$,  has  the advantage of the implementation simplicity (scaling only) and a relatively simple design (one parameter needs to be found). Then, the main question is  how to choose the correction factor $\alpha$.

The contributions in \cite{Lechner07}\cite{Alvarado09b} attempted to answer this question making $f(\alpha)=\alpha\cd l$ ``close'' to $f^\tr{c}(l)$. In particular, \cite{Lechner07} find the correction factor via the weighted least-square fit (WLSF) to the function $f^\tr{c}(l)$
\begin{align}
\hat{\alpha}^\tr{WLSF}&=\tr{argmin}_{\alpha}\ms{E}_{\tilde{L}_{n}|C_{n}=0}\bset{|f^\tr{c}(\tilde{L}_{n})-\alpha\tilde{L}_{n}|^{2}}\nonumber\\
&=\tr{argmin}_{\alpha}\int_{-\infty}^{\infty}p_{\tilde{L}_{n}|C_{n}}(l|0)\bigl(f^\tr{c}(l)-\alpha\cd l\bigr)^{2}\tr{d}l.\label{WLSF}
\end{align}
This criterion, however, is not associated with the performance of the decoder. Moreover, since we use the function $f^\tr{c}(l)$, the form of the pdf $p_{\tilde{L}_{n}|C_{n}}(l|0)$ has to be known or explicitly estimated.

In the recent works \cite{Nguyen11}\cite{Yazdani11}, the correction factor was found through maximization of the generalized mutual information (GMI) \cite{Martinez09} between the mismatched L-values and the corresponding bits. Assuming \eqref{symm.cond}, this criterion boils down to solving the following optimization problem
\begin{align}
\hat{\alpha}^\tr{GMI}&=\tr{argmin}_{\alpha}\ms{E}_{\tilde{L}_{n}|C_{n}=0}\bset{\log_{2}(1+\tr{e}^{\tilde{L}\alpha})}\nonumber\\
&=\tr{argmin}_{\alpha}\int_{-\infty}^{\infty}p_{\tilde{L}_{n}|C_{n}}(l|0)\log_{2}(1+\tr{e}^{l\alpha})\tr{d}l.\label{GMI}
\end{align}

The minimum is reached when the derivative of the integral in \eqref{GMI} goes to zero
\begin{align}\label{}
\frac{\tr{d}}{\tr{d}\alpha}\ms{E}_{\tilde{L}_{n}|C_{n}=0}\set{\log(1+e^{\alpha \tilde{L}_{n}})}&=0\nonumber\\
\int_{-\infty}^{\infty}p_{\tilde{L}_{n}|C_{n}}(l|0)\frac{l\cd e^{\alpha l}}{1+e^{\alpha l}}\tr{d}l &=0\label{cond.GMI.2}\\
\int_{-\infty}^{\infty}p_{\tilde{L}_{n}|C_{n}}(l|0)\frac{l\cd e^{\alpha \frac{l}{2}}}{\cosh(\alpha \frac{l}{2})}\tr{d}l &=0\label{cond.GMI.3}.
\end{align}

While it is argued (and demonstrated on examples) in \cite{Nguyen11}\cite{Yazdani11} that the maximization of GMI should improve the performance of the capacity-approaching codes, the correction criterion \eqref{GMI} does not relate directly to the performance of the ML decoder we are interested in. 

Moreover, solving \eqref{GMI} requires the numerical quadratures as the the logarithm within \eqref{GMI} or the hyperbolic cosine within \eqref{cond.GMI.3} will resist analytical integration.  If $p_{\tilde{L}_{n}|C_{n}}(l|0)$ is not known, the Monte-Carlo integration (simulations) may be used to calculate the integral in \eqref{GMI} or \eqref{cond.GMI.3} but such an approach precludes the on-line (i.e., model-based) correction. However, it is still simpler to implement than \eqref{WLSF} because we do not need to know the function $f^\tr{c}(l)$.

\subsection{PEP}\label{Sec:PEP}
To describe the behaviour of the ML decoder \eqref{ML.dec.LLR} based on the corrected L-values
\begin{align}
\hat{\bc}&=\tr{argmax}_{\bc\in\mc{C}}\sum_{n=1}^{N} \alpha_{n}\cd\tilde{l}_{n}\cd c_{n},\label{ML.dec.mm}
\end{align}
we will use  the pairwise error probability (PEP) defined as the probability of detecting codeword $\hat{\bc}$ when sending the codeword $\bc$.

Assuming that the code $\mc{C}$ is linear and \eqref{symm.cond} holds, instead of calculating the PEP for all pairs ($\bc, \hat{\bc}$) it is enough to calculated the PEP for all $\hat{\bc}\neq \bc$ assuming the all-zeros codeword $\bc=[0,\ld, 0]$ was sent, that is, the probability of the event $\bc\rightarrow\hat{\bc}$
\begin{align}
\tr{PEP}(\set{\alpha_{n}}_{n=1}^{N},\hat{\bc})
&=\tr{Pr}\bset{\sum_{n=1}^{N}\alpha_{n}\cd\tilde{L}_{n}\cd\hat{c}_{n}>\alpha_{n}\cd\tilde{L}_{n}\cd c_{n}}\label{PEP.a}\\
&=\tr{Pr}\bset{\sum_{n=1}^{N}\tilde{L}^\tr{c}_{n}\cd\hat{c}_{n}>0}\label{PEP}\\
&=\int_0^{\infty}  \bigl[p_{\tilde{L}^\tr{c}_{1}|C_{1}}(l|0)\bigr]^{\hat{c}_{1}}\star\ld\star \bigl[p_{\tilde{L}^\tr{c}_{N}|C_{N}}(l|0)\bigr]^{\hat{c}_{N}} \tr{d} l\label{PEP.d.int}
\end{align}
where $\star$ is the convolution operator.

This notation emphasizes that the PEP depends on the correction factors $\set{\alpha_{n}}_{n=1}^{N}$ and the codeword $\hat{\bc}$.

If we denote by  $\set{n_{k}}_{k=1}^{d}$, the set of indices such that $\hat{c}_{n_{k}}=1$, where $d$ is the Hamming weight of $\hat{\bc}$, the PEP \eqref{PEP.d.int}  can be written as 
\begin{align}
\tr{PEP}(\set{\alpha_{n}}_{n=1}^{N},\hat{\bc})
&=\int_0^{\infty} p_{\tilde{L}^\tr{c}_{n_{1}}|C_{n_{1}}}(l|0)\star\ld\star p_{\tilde{L}^\tr{c}_{n_{d}}|C_{n_{d}}}(l|0) \tr{d} l.\label{PEP.d.int.2},
\end{align}
that is, it depends solely on the pdfs of the L-values indexed by $\set{n_{k}}_{k=1}^{d}$.

We also note quickly that multiplying all the L-values $\tilde{l}_{n}$ in \eqref{ML.dec.mm} by $\alpha_{n}\equiv\alpha$ cannot change the decoding results so,  in such a case, the linear correction is useless if ML decoder is used. However, it still may be useful if another type of decoding is applied. For example, iterative decoders (e.g., of the turbo codes or LDPC codes) may benefit from such a correction.

\section{PEP-minimizing correction}\label{Sec:Min.PEP}
Now, we want answer the question: how to choose the correction factors $\set{\alpha_{n}}_{n=1}^{N}$ so that the error of the decoder that uses the corrected L-values $\tilde{l}^\tr{c}_{n}=\alpha_{n}\cd\tilde{l}_{n}$ is minimized? 

From the previous discussion we conclude that, in order to improve the performance of the decoder,  we should find  $\set{\alpha_{n}}_{n=1}^{N}$ to minimizes the $\tr{PEP}\left(\set{\alpha_{n}}_{n=1}^{N}, \hat{\bc}\right)$ in \eqref{PEP.d.int} for \emph{any} codeword $\hat{\bc}$. Thus, we have to solve the following optimization problem
 \begin{align}\label{opt.alpha}
\set{\hat{\alpha}_{n}}_{n=1}^{N}=\tr{argmin}_{\set{\alpha_{n}}_{n=1}^{N}}\tr{PEP}\Bigl(\set{\alpha_{n}}_{n=1}^{N}; \hat{\bc}\Bigr),
\end{align}
and its solution should be independent of $\hat{\bc}$ because we want to apply the correction factors to all L-values prior to decoding and we do not know which error ($\bc\rightarrow\hat{\bc}$) will occur.

At first sight, the problem may appear intractable due to the dependance of the PEP on various $\hat{\bc}$, each resulting in a different set of L-values indexed by $\set{n_{k}}_{k=1}^{d}$ which are then convolved as per \eqref{PEP.d.int.2}.

\subsection{Two-state mismatch}\label{Sec:Binary.Mismatch}
Before attacking the problem \eqref{opt.alpha} we will analyze a simpler case of a \emph{two-state mismatch}, where $N_{1}$ of the L-values are independent identically distributed (i.i.d) and mismatched and the remaining $N-N_{1}$ are i.i.d and matched. In this case, all the mismatched L-values will be multiplied by the same correction factor $\alpha$, that is, $\alpha_{n}=\alpha, n=1, \ld, N_{1}$ and the matched L-values will remain unaltered.

Since, we do not known a priori the indices $\set{n_{k}}_{k=1}^{d}$, we do not know a priori how many mismatched metrics will affect the PEP.  We thus assume initially that among the L-values affecting the PEP calculation, $d_{1}$ are mismatched and $d_{2}$ L-values are matched.  This specifies \eqref{PEP.d.int.2} as follows
\begin{align}
\tr{PEP}\Bigl(\set{\alpha_{n}}_{n=1}^{N}\Bigr)&=\tr{PEP}(\alpha)
=\tr{Pr}\bset{\sum_{k=1}^{d_{1}}\alpha\cd\tilde{L}_{k}+\sum_{k=1}^{d_{2}}L_{k}<0}\nonumber\\
&=\int_0^{\infty} \Bigl[p_{\tilde{L}^\tr{c}|C}({l}|0)\Bigr]^{\star d_{1}} \star \Bigl[p_{L|C}(l|0)\Bigr]^{\star d_{2}} \tr{d} l,\label{PEP.d.int.mismatch}
\end{align}
where  $[f(l)]^{\star d}$ is a $d$-fold self-convolution of $f(l)$ and we emphasize that the PEP depends uniquely on one parameter $\alpha$.

We want minimize $\tr{PEP}(\alpha)$ \eqref{PEP.d.int.mismatch} for any $d_{1}$ and $d_{2}$ thus, the solution of 
\begin{align}\label{solve.alpha}
\hat{\alpha}=\tr{argmin}_{\alpha}\tr{PEP}(\alpha)
\end{align}
should be independent of $d_{1}$ and $d_{2}$.

\begin{example}\label{Ex.1}
Assume that the bits $c_{n}$ are sent using a binary phase-shift keying (BPSK) modulation so 
\begin{align}\label{p.y}
p_{Y_{n}|C_{n}}(y_{n}|c_{n})=\Psi\left(y_{n}-(2c_{n}-1); \frac{1}{2\gamma}\right),
\end{align}
where  $\gamma$ has the meaning of the average signal-to-noise ratio (SNR) and $\Psi(\cd)$ is given by \eqref{Gauss.function}.

To calculate the L-values via \eqref{LLR.intro} using \eqref{p.y} we need to know the value of $\gamma$ and we assume that its estimate $\tilde{\gamma}\neq\gamma$ is used for the first $N_{1}$ L-value as $\tilde{l}_{n}=4y_{n}\cd\tilde{\gamma}, l=1,\ld, N_{1}$ so these L-values are mismatched. The exact value of $\gamma$ is used for the remaining L-values $l_{n}=4y_{n}\cd\gamma, l=N_{1}+1,\ld, N$ and these L-value are matched. It is straightforward to see that  the pdf of the mismatched L-values is given by $p_{\tilde{L}_{n}|C_{n}}(l | 0)= \Psi(  l + 4\tilde{\gamma}; 8\tilde{\gamma}^{2}/\gamma) $ while the pdf of the matched L-values by  $p_{L_{n}|C_{n}}(l | 0)= \Psi(  l + 4\gamma; 8\gamma) $ \cite{Martinez07}.

Since all the L-values affecting the PEP are Gaussian, the result of their convolution is also Gaussian and we can write \eqref{PEP.d.int.mismatch} as
\begin{align}\label{PEP.2Gauss}
  \tr{PEP}(\alpha)&=Q\left(\frac{\alpha d_{1}\tilde{\mu}+d_{2}\mu}{\sqrt{\alpha^{2} d_{1}\tilde{\sigma}^{2}+d_{2}\sigma^{2}}}\right)
  =Q\left(\sqrt{2}\frac{\alpha d_{1}\tilde{\gamma}+d_{2}\gamma}{\sqrt{\alpha^{2}d_{1}\tilde{\gamma}^{2}/\gamma+d_{2}\gamma}}\right),
  \end{align}
where $Q(x)=\frac{1}{\sqrt{2\pi}}\int_{x}^{\infty}\exp(-t^{2}/2)\tr{d}t$.

Verifying that \eqref{PEP.2Gauss} is convex with respect to $\alpha$ and setting its derivative to zero yields  the global minimum of  \eqref{solve.alpha} given by  $\hat{\alpha}=\gamma/\tilde{\gamma}$. 

Note that, as required, the correction factor is independent of $d_{1}$ and $d_{2}$ thus the PEP is minimized independently of the error event $\bc\rightarrow\hat{\bc}$.

We can also immediately see that $\tilde{l}^\tr{c}_{n}=\alpha\cd \tilde{l}_{n}=\tilde{l}_{n}\gamma/\tilde{\gamma}=4y_{n}\gamma$, that is, $\tilde{l}^\tr{c}_{n}=\l_{n}$ and we recover the matched L-value. Of course, if we knew that $\gamma$ should be used, we would not use $\tilde{\gamma}$ to calculate the L-values, in the first place so this example illustrates only the principle of correction.
\end{example}

\subsection{Approximation of the PEP}
To apply the PEP-minimization principle \eqref{solve.alpha} in a general case, we must be able to find the PEP for arbitrary distributions of the L-values. Since, in general,  this cannot be done exactly in an analytical form, we will turn to approximations. 

Defining 
\begin{align}\label{}
L^{\Sigma}=\sum_{k=1}^{d_{1}}\tilde{L}^\tr{c}_{k}+\sum_{k=1}^{d_{2}}L_{k}=\sum_{k=1}^{d_{1}}\alpha\tilde{L}_{k}+\sum_{k=1}^{d_{2}}L_{k}
\end{align}
we can write \eqref{PEP.d.int.mismatch} as $\tr{PEP}(\alpha)=\tr{Pr}\bset{L^{\Sigma} > 0}$. 

Then, the Bhattacharyya upper bound for the PEP is given by \cite{Caire98}\cite{Martinez06}
\begin{align}\label{B-UB}
\tr{PEP}(\alpha)\leq \tr{PEP}^\tr{UB}(\alpha)=\tr{e}^{\kappa_{\Sigma}(\hat{s})},
\end{align}
where $\kappa_{\Sigma}(s)$ is the cumulant generating function (CGF) of $L^{\Sigma}$ 
\begin{align}\label{}
\kappa_{\Sigma}(s)=d_{1}\kappa_{\tilde{L}^\tr{c}}(s )+d_{2}\kappa_{L}(s)=d_{1}\kappa_{\tilde{L}}(s \alpha)+d_{2}\kappa_{L}(s),
\end{align}
and
\begin{align}\label{}
\kappa_{L}(s)&=\log\ms{E}_{L|C=0}\set{\tr{e}^{sL}}=\log\int_{-\infty}^{\infty}p_{{L}|0}(l)\tr{e}^{sl}\tr{d}l \\
\kappa_{\tilde{L}}(s)&=\log\ms{E}_{\tilde{L}|C=0}\set{\tr{e}^{s\tilde{L}}}\\
\kappa_{\tilde{L}^\tr{c}}(s)&=\log\ms{E}_{\tilde{L}|C=0}\set{\tr{e}^{s\alpha \tilde{L}}}=\kappa_{\tilde{L}}(s\alpha).
\end{align}

In \eqref{B-UB}, $\hat{s}=\tr{argmin}_{s\in\mc{R}}\kappa_{\Sigma}(s)$ is the so-called saddlepoint \cite{Martinez06} of $\kappa_{\Sigma}(s)$, which is unique because the CGF is always convex.

The bound \eqref{B-UB} was shown in \cite{Martinez06} to be quite loose and a much more accurate estimation of the PEP can be obtained using the so-called saddlepoint approximation (SPA) \cite{Martinez06}\cite{Szczecinski09b}\cite{Kenarsari10}
\begin{align}\label{PEP.SPA}
   \tr{PEP}(\alpha)\approx \widetilde{\tr{PEP}}(\alpha)=\frac{\tr{e}^{\kappa_{\Sigma}(\hat{s})}}{|\hat{s}|\sqrt{2\pi \kappa''_{\Sigma}(\hat{s})}}.
\end{align}

However, minimization of \eqref{PEP.SPA} is quite difficult due to the implicit dependence of $\hat{s}$ on $\alpha$, therefore, for simplicity we opt for minimization of the upper bound  \eqref{B-UB}. Nevertheless, even if the correction factors $\alpha$ minimizing of $\tr{PEP}^\tr{UB}(\alpha)$ and $\widetilde{\tr{PEP}}(\alpha)$ would not be the same, we expect them to be similar as the exponential term $\tr{e}^{\kappa_{\Sigma}(s)}$ dominates both expressions.

\begin{theorem}\label{Lemma1}
The  upper bound for the PEP in \eqref{B-UB} is minimized setting the correction factor to $\hat{\alpha}=\frac{\hat{s}_{1}}{\hat{s}_{2}}$, where $\hat{s}_{1}$ and $\hat{s}_{2}$ are the saddlepoints of the matched and mismatched L-values, that is, $\kappa'_{\tilde{L}}(\hat{s}_{1})=0$ and $\kappa'_{L}(\hat{s}_{2})=0$. 
\begin{IEEEproof}
We start bounding \eqref{B-UB} as 
\begin{align}
\min_{\alpha}\tr{PEP}^\tr{UB}(\alpha)
&=\tr{e}^{\min_{\alpha,s}\kappa_{\Sigma}(s)}=\tr{e}^{\min_{\alpha,s}\bigl(d_{1}\kappa_{\tilde{L}}(s\alpha) + d_{2}\kappa_{L}(s) \bigr)   }\label{PEP.bound.1}\\
&\geq\tr{e}^{ d_{1}\kappa_{\tilde{L}}(\hat{s}_{1})  + d_{2}\kappa_{L}(\hat{s}_{2})   }\label{PEP.bound},
\end{align}
where \eqref{PEP.bound} is the global minimum of \eqref{B-UB}.

We can see that the exponent in \eqref{PEP.bound.1} $\min_{\alpha}\bigl(d_{1}\kappa_{\tilde{L}}(s\alpha) + d_{2}\kappa_{L}(s) \bigr)$ reaches its global minimum for $s\alpha=\hat{s}_{1}$ and $s=\hat{s}_{2}$, that is, when  $\alpha=\hat{s}_{1}/\hat{s}_{2}$; this means that the saddle point of $\kappa_{\Sigma}(s)$ is $\hat{s}=\hat{s}_{2}$. 
\end{IEEEproof}
\end{theorem}

As required, the bound on the PEP is minimized independently of $d_{1}$ and $d_{2}$.

\subsection{Arbitrary mismatch}

We are now ready to abandon the context of the two-state mismatch and may  extend the previous result to the case treated in \eqref{opt.alpha}.

Let $L^{\Sigma}=\sum_{n=1}^{N}\alpha_{n}\cd \tilde{L}_{n}\cd \hat{c}_{n}$ has the CGF given by $\kappa_{\Sigma}(s)=\sum_{n=1}^{N}\kappa_{\tilde{L}_{n}}(s\cd \alpha_{n})\cd\hat{c}_{n}$, where $\kappa_{n}(s)$ is the CGF of the L-value $\tilde{L}_{n}$ conditioned on $C_{n}=0$. Define the upper bound on the PEP \eqref{opt.alpha} as
\begin{align}\label{PEP.SPA.general}
\tr{PEP}\bigl(\set{\alpha_{n}}_{n=1}^{N};\hat{\bc}\bigr)\leq \tr{PEP}^\tr{UB}\bigl(\set{\alpha_{n}}_{n=1}^{N}; \hat{\bc}\bigr)=\tr{e}^{\kappa_{\Sigma}(\hat{s})}.
\end{align}

\begin{proposition}
The linear correction factors that minimize the upper bound on PEP \eqref{PEP.SPA.general} are given by $\hat{\alpha}_{n}=\frac{\hat{s}_{n}}{\hat{s}_{0}}$, where $\hat{s}_{n}$ is the saddlepoint of the CGF $\kappa_{n}(s)$, and $\hat{s}_{0}>0$ is chosen arbitrarily. 
\begin{proof} As in the proof  of Theorem~\ref{Lemma1} we write
\begin{align}
\min_{\set{\alpha_{n}}_{n=1}^{N}}\tr{PEP}^\tr{UB}\bigl(\set{\alpha_{n}}_{n=1}^{N}; \hat{\bc}\bigr)
&=\tr{e}^{\min_{s,\set{\alpha_{n}}_{n=1}^{N}}\bigl( \sum_{n=1}^{N}\kappa_{\tilde{L}_{n}}(s\cd \alpha_{n})\cd\hat{c}_{n} \bigr)   }\label{PEP.bound.2}\\
&\geq\tr{e}^{ \sum_{n=1}^{N}\kappa_{\tilde{L}_{n}}(\hat{s}_{n})\cd\hat{c}_{n}   }.
\end{align}
The global minimum is reached when $\hat{s}_{n}=s\alpha_{n}, n=1,\ld, N$. This produces $N$ equations with $N+1$ variables and since $s$ and $\hat{s}_{1},\ld, \hat{s}_{N}$ are positive we may arbitrarily fix $s=\hat{s}_{0}$.
\end{proof}
\end{proposition}

\textbf{Remark~1}:~Although $\hat{s}_{0}$ may be chosen arbitrarily (recall that the multiplication of all L-values by the same correction factor does not change the ML decoding results), it is reasonable to use $\hat{s}_{0}=\frac{1}{2}$. This is because the saddlepoint of the matched metrics equals $\hat{s}_{n}=\frac{1}{2}$ \cite{Martinez06} and their correction factor is then given by $\hat{\alpha}=1$, that is, no correction is necessary as we would expect it. Thus, the simple rule consists  in doubling the saddlepoint of the L-values' CGF
\begin{align}\label{corr.factor}
  \hat{\alpha}_{n}=2\hat{s}_{n}.
\end{align}

\textbf{Remark~2}: We recall that if we want to use the pdf conditioned on $C_{n}=1$, $p_{\tilde{L}_{n}|C_{n}}(l|1)$, instead of   $p_{\tilde{L}_{n}|C_{n}}(l|0)$, the saddlepoint in negative $\hat{s}_{n}<0$, but then also for the matched metrics $\hat{s}_{0}=-\frac{1}{2}$. Thus, to take both cases into account we might reformulate \eqref{corr.factor} as
\begin{align}\label{corr.factor.2}
  \hat{\alpha}_{n}=2|\hat{s}_{n}|.
\end{align}

\textbf{Remark~3}: Since the CGF of the L-values $\tilde{L}^\tr{c}_{n}$ (after correction) is equal to $\kappa_{\tilde{L}^\tr{c}_{n}}(s)=\kappa_{\tilde{L}_{n}}(s\alpha_{n})$, the saddlepoint of $\kappa_{\tilde{L}^\tr{c}_{n}}(s)$ is given by $\hat{s}_{n}/\hat{\alpha}_{n}=\hat{s}_{0}$. That is, the saddlepoint of the CGF of each corrected L-values is equal to $\hat{s}_{0}=\frac{1}{2}$.

\subsection{Relationship to the GMI-maximizing correction}
Let us compare now the correction factor defined using \eqref{corr.factor} to $\hat{\alpha}^\tr{GMI}$ defined in \eqref{cond.GMI.3} where the maximization is obtained finding the zero of its derivative
\begin{align}\label{}
\int_{-\infty}^{\infty}p_{\tilde{L}|C}(l|0)\frac{l\cd \tr{e}^{\alpha \frac{l}{2}}}{\cosh(\alpha \frac{l}{2})}\tr{d}l &=0\label{cond.GMI.4}\\
\int_{-\infty}^{\infty}p_{\tilde{L}^\tr{c}|C}(l |0)\tr{e}^{\frac{l}{2}}\frac{l}{\cosh(\frac{l}{2})}~\tr{d}l &=0\label{cond.GMI.5},
\end{align}
where \eqref{cond.GMI.5} is obtained from \eqref{cond.GMI.4} after the change of variables using the pdf of the corrected L-value $\tilde{L}^\tr{c}=\alpha \tilde{L}$, i.e.,  $p_{\tilde{L}^\tr{c}|C}(l|0)=\alpha^{-1}\cd p_{\tilde{L}^\tr{c}|0}(\alpha^{-1}\cd l|0)$.

On the other hand, the condition we derived in \eqref{corr.factor} states that the saddlepoint of $\kappa_{\tilde{L}^{c}}(s)$ should be equal to $\frac{1}{2}$, which may be written as follows
\begin{align}\label{}
\frac{\tr{d}}{\tr{d}s}\kappa_{\tilde{L}^\tr{c}}(s)|_{s=\frac{1}{2}}=\frac{\ms{E}_{\tilde{L}^\tr{c}|C=0}\set{ \tilde{L}^\tr{c}\tr{e}^{\frac{\tilde{L}^\tr{c}}{2}}}}{\ms{E}_{\tilde{L}^\tr{c}|C=0}\set{\tr{e}^{\frac{\tilde{L}^\tr{c}}{2}}}}&=0\\
\int_{-\infty}^{\infty}p_{\tilde{L}^\tr{c}|C}(l|0) e^{\frac{l}{2}}\cd l~\tr{d}l &=0.\label{cond.SP}
\end{align}

Since $\cosh(\frac{l}{2})$ in the denominator of \eqref{cond.GMI.5} is symmetric, we can see that if $p_{\tilde{L}^{c}|C}(l|0)\tr{e}^{\frac{l}{2}}$ is symmetric, both \eqref{cond.GMI.5} and \eqref{cond.SP} are satisfied. This is also the condition we derived in \eqref{cons.symm.cond} and can be interpreted as follows:  if the optimal correction function $f^\tr{c}(l)$ is linear, i.e., after the linear correction the L-value satisfies the symmetry-consistency condition \eqref{cons.symm.cond}, both criteria yield the same correction factor. In general, however, they need not be the same.

\section{Example of application: Correcting the interference effects}\label{Sec:Num.Results}

Consider a BPSK transmission, when the sent symbols $x_{n}=2c_{n}-1$ pass through a channel affected by additive Gaussian noise and a BPSK interference
\begin{align}\label{}
y_{n}=h_{n}\cd x_{n} + z_{n} + g_{n}\cd d_{n},
\end{align}
where $h_{n}\in\mc{R}$ is the known gain of the channel, $z_{n}$ is a zero-mean Gaussian signal with known variance $\sigma_{z}^{2}=N_{0}/2$; $d_{n}\in\set{-1, 1}$ is the BPSK-modulated interference signal received with the gain $g_{n}\in\mc{R}$. We define also the SNR and the signal-to-interference ratio (SIR), respectively, as $SNR=h_{n}^{2}/N_{0}$ and $SIR=h_{n}^{2}/g_{n}^{2}$. 

Although using \eqref{LLR.intro}  it is relatively simple to calculate the L-values in this case as 
\begin{align}\label{L.value-sumexp}
l_{n}=\log\frac{\exp(-\frac{(y_{n}-h_{n}-g_{n})^{2}}{2\sigma_{z}^{2}})+\exp(-\frac{(y_{n}-h_{n}+g_{n})^{2}}{2\sigma_{z}^{2}})}
    {\exp(-\frac{(y_{n}+h_{n}-g_{n})^{2}}{2\sigma_{z}^{2}})+\exp(-\frac{(y_{n}+h_{n}+g_{n})^{2}}{2\sigma_{z}^{2}})},
\end{align}
for the purpose of our examples, the receiver ignores the interference, thus assumes $g_{n}=0$, and then, from \eqref{L.value-sumexp} we obtain the L-value
\begin{align}\label{l.tilde.BPSK}
  \tilde{l}_{n}=\frac{2 h_{n}\cd y_{n}}{\sigma_z^{2}},
\end{align}
which is mismatched due to assumed absence of interference.

To apply the correction principle we derived, we need to calculate the CGF of $\tilde{L}=\frac{2 h_{n}}{\sigma_z^{2}}\cd Y$, conditionned on sending the bit $c_{n}=0$, $\kappa_{\tilde{L}}(s)=\kappa_{Y}(\frac{2h_{n}}{\sigma_z^{2}} s)$, where 
\begin{align}\label{kappa.Y}
\kappa_{Y}(s)=-h_{n}s+\frac{1}{2}\sigma_{z}^{2}s^{2}+ \log\left(\frac{1}{2}\tr{e}^{s\cd g_{n}}+\frac{1}{2}\tr{e}^{-s\cd g_{n}}\right).
\end{align}
Then, finding the saddlepoint of $\kappa_{Y}(s)$, as $\kappa_{Y}'(\hat{s})=0$, the saddlepoint of $\kappa_{\tilde{L}}(s)$ equals to $\frac{\sigma^{2}_{z}}{2h_{n}}\hat{s} $ thus, the correction factor applied to the $\tilde{l}_{n}$ is given by \eqref{corr.factor} $\hat{\alpha}=\frac{\sigma_{z}^{2}}{h_{n}}\hat{s}$ and the correction boils down to the calculation 
\begin{align}\label{}
\tilde{l}^{\tr{c}}_{n}=2\hat{s}\cd y_{n}.
\end{align}

Note that, instead of calculating the saddlepoint of $\kappa_{\tilde{L}}(s)$ and correcting $\tilde{l}_{n}$, we might directly calculate the saddlepoint of $\kappa_{Y}(s)$ and apply it to the observation $y_{n}$. This is possible, of course, because the L-value $\tilde{l}_{n}$ is a scaled  version of $y_{n}$, cf.~\eqref{l.tilde.BPSK}.

\begin{figure}[t]
\psfrag{LHS}[l][l]{LHS of \eqref{SP.eq.r}}
\psfrag{RHS}[c][l]{RHS of \eqref{SP.eq.r}}
\psfrag{ZZx}[c][c]{0}
\psfrag{gn}[c][c]{$g_{n}$}
\psfrag{hn}[c][c]{$h_{n}$}
\psfrag{hnsig2z}[c][c]{$\frac{h_{n}}{\sigma^{2}_{z}}$}
\psfrag{shat}[t][c]{$\hat{s}$}
\psfrag{s0}[t][c]{$\hat{s}^{0}$}
\psfrag{sinf}[t][r]{$\hat{s}^{\infty}$}
\begin{center}
\scalebox{0.9}{\includegraphics[width=1\linewidth]{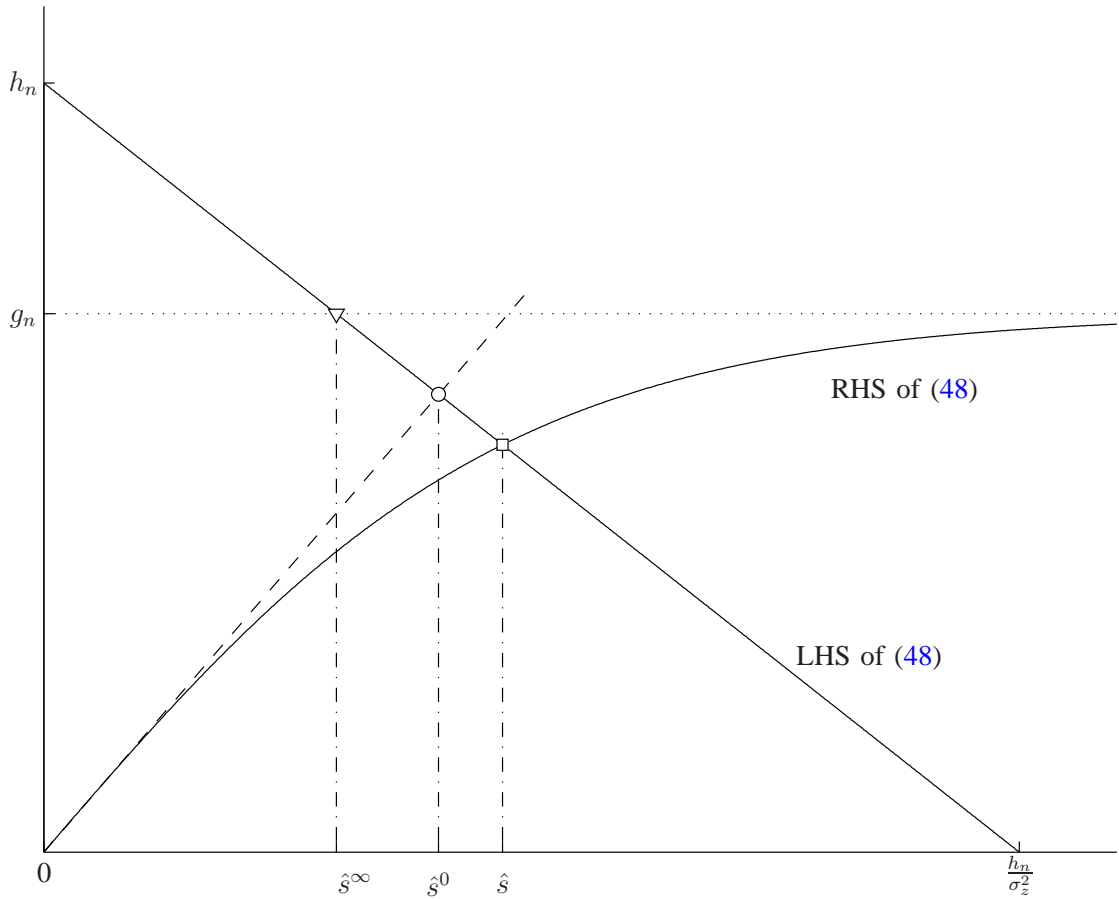}}

\caption{Graphical interpretation of the solution of the saddlepoint equation \eqref{SP.eq.r}; the dashed line correspond to the linearization  $\tanh(x)\approx x$.}\label{Fig:GRSol}
\end{center}
\end{figure}

To find the saddlepoint $\hat{s}$ we differentiate \eqref{kappa.Y} to obtain the following saddlepoint equation
\begin{align}\label{SP.eq.r}
h_{n}  - \sigma_{z}^{2}\cd\hat{s} = g_{n}\cd\tanh( g_{n}\cd \hat{s} ),
\end{align}
whose graphical interpretation as the intersection of the right-had- and left-hand sides is shown in Fig.~\ref{Fig:GRSol}. 

While  \eqref{SP.eq.r} cannot be solved in a closed-form, we may obtain approximations in particular cases. Namely, for  $SNR\rightarrow 0$ (i.e., when $\sigma_{z}^{2}\rightarrow \infty)$, we easily see that $\hat{s}\rightarrow 0$ so, using the linearization $\tanh(x)\approx x$ (shown as a dashed line in Fig.~\ref{Fig:GRSol}), we obtain
\begin{align}\label{}
\hat{s}^{0}=\frac{h_{n}}{\sigma_{z}^{2}+g_{n}^{2}}=\frac{h_{n}}{\sigma_{\tr{N+I},n}^{2}}
\end{align}
where $\sigma_{\tr{N+I},n}^{2}$ is the variance of the noise and interference. The corresponding correction factor is given by $\alpha^{0}=\frac{\sigma_{z}^{2}}{\sigma_{\tr{N+I},n}^{2}}$. Note that using $\tilde{\mu}=-\ms{E}_{\tilde{L}_{n}|C_{n}=0}=\frac{h^{2}_{n}}{\sigma^{2}_{z}}$ and $\tilde{\sigma}^{2}=\tr{Var}\set{\tilde{L}_{n}}=\frac{4h^{2}_{n}}{\sigma^{4}_{z}}$ in \eqref{alpha.Gauss} yields exactly the same results $\alpha^\tr{Gauss}=\alpha^{0}$. 

This means that for the low SNR, when the noise ``dominates'' the interference, the effect of noise and the interference can be modelled as a Gaussian variable with variance $\sigma_{\tr{N+I},n}^{2}$; this is often done in practice \cite{Sellathurai02} and the corrected L-values, in this case, would be calculated as
\begin{align}\label{Lcorr.0}
\tilde{l}^{\tr{c},0}_{n}=\frac{2h_{n}\cd y_{n}}{\sigma_{\tr{N+I},n}^{2}}.
\end{align}

\begin{figure}[t]
\psfrag{xlabel}[c][c]{$SNR$}
\psfrag{ylabel}[c][c]{}
\psfrag{XXX1}{$\hat{\alpha}^\tr{GMI}$}
\psfrag{XXX2}{$\hat{\alpha}$}
\psfrag{XXX3}{$\hat{\alpha}^\tr{2SM}$}
\psfrag{XXX4}{$\hat{\alpha}^\tr{WLSF}$}
\psfrag{SIR3}[l][c]{$SIR=3$~dB}
\psfrag{SIR6}[c][c]{$SIR=6$~dB}
\psfrag{SIR12}[c][c]{$SIR=12$~dB}
\psfrag{SIR10}[c][c]{$SIR=10$~dB}
\begin{center}
\scalebox{0.85}{\includegraphics[width=1\linewidth]{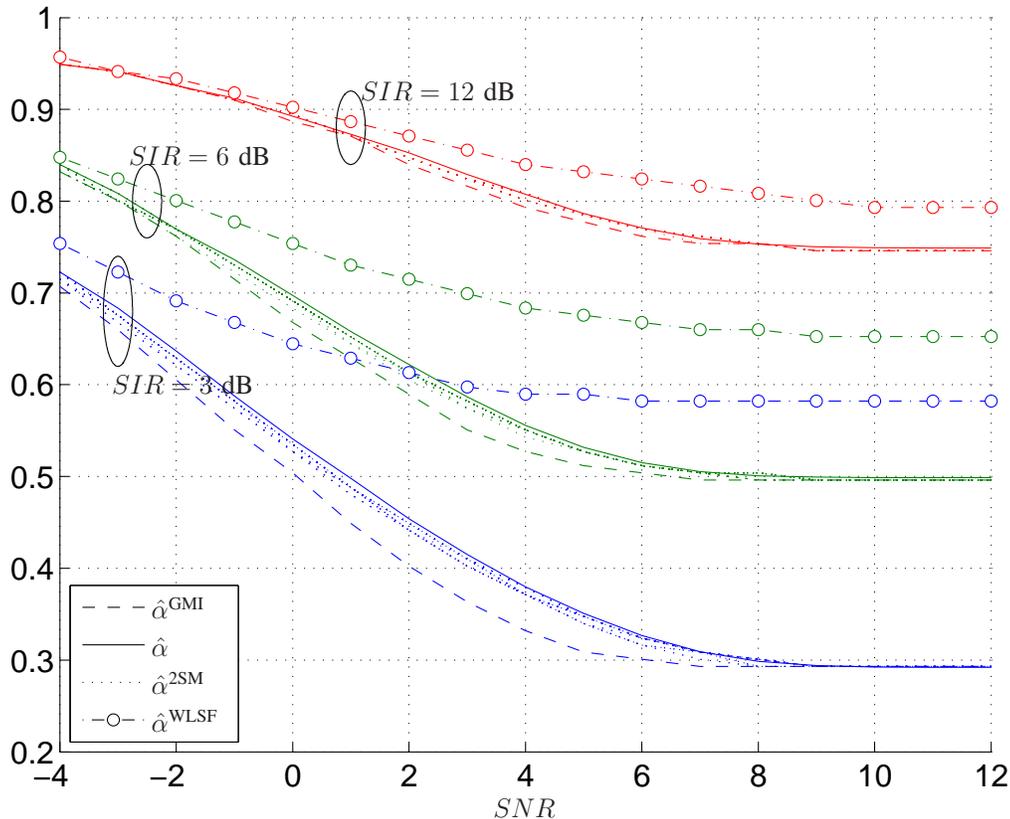}}

\caption{Correction factors  $\hat{\alpha}^\tr{GMI}$, $\hat{\alpha}$ $\hat{\alpha}^\tr{WLSF}$obtained for various values of SNR and SIR. The dotted line show the value of $\hat{\alpha}^\tr{2SM}$, cf.~\eqref{alpha.PEP} for various combinations of $2\leq d_{1},d_{2}\leq 8$. }\label{Fig:Corr.Fact}
\end{center}
\end{figure}

For high SNR, i.e., when $SNR\rightarrow\infty$ (i.e., $\sigma_{z}^{2}\rightarrow 0$), we observe that $\hat{s}\rightarrow-\infty$ so taking advantage in \eqref{SP.eq.r}  of the saturation of $\tanh(\infty)= 1$, the saddlepoint is given by\footnote{We assume $g_{n}<h_{n}$, i.e., the interference is weaker than the desired signal.}
\begin{align}\label{}
 \hat{s}^{\infty} =\frac{h_{n}-g_{n}}{\sigma_{z}^{2}},
\end{align}
and the corresponding correction factor by
\begin{align}\label{alpha.inf}
\hat{\alpha}^{\infty}=1-g_{n}/h_{n}.
\end{align}

The L-values in this case would be calculated as
\begin{align}\label{l.tilde.BPSK.inf}
\tilde{l}^{\tr{c},\infty}_{n}=\frac{2 (h_{n}-g_{n})\cd y_{n}}{\sigma_{z}^{2}},
\end{align}
that is, the interference-related term decreases the gain of the useful signal. Intuitively, this can be explained as follows: for high SNR, the interference can be ``distinguished'' from the noise and becomes the part of the transmitted constellation, i.e., sending bit $c_{n}=1$, we will effectively be able to make a difference between $h_{n}-g_{n}$ or $h_{n}+g_{n}$. Moreover, for high SNR the symbol that is the most likely to provoke the error is the one closest to the origin, that is $h_{n}-g_{n}$. This leads to assuming that BPSK symbols are sent over a channel with gain $h_{n}-g_{n}$, cf.~\eqref{l.tilde.BPSK.inf}.

For $0<SNR<\infty$, the saddlepoint can be obtained numerically solving \eqref{SP.eq.r}. For example, we might use the Newton-Raphson method
\begin{align}\label{Newton.iterations}
\hat{s}(i)=\hat{s}(i-1)-\frac{\kappa'_{Y}\bigl(\hat{s}(i-1)\bigr)}{\kappa''_{Y}\bigl(\hat{s}(i-1)\bigr)}, \quad i=1, 2, \ld, I_\tr{max}
\end{align}
where 
\begin{align}\label{kappa.bis}
\kappa_{Y}''(s)=\sigma_{z}^{2} + \frac{g^{2}}{\cosh^{2}(g\cd s)},
\end{align}
and $\hat{s}(0)$ is  the appropriately chosen starting point for the recursion, e.g., $\hat{s}(0)=\max\set{\hat{s}^{\infty},\hat{s}^{0}}$. In this work, we used $I_\tr{max}=2$ so a small computation load in incurred due to the on-line calculation of the the correction factors; alternatively, the correction factors might be interpolated using a table precalculated for different values of $SNR$ and $SIR$.

Since $\frac{h_{n}}{\sigma_{z}^{2}}>\hat{s}$ (cf.~Fig.~\ref{Fig:GRSol}) we can also immediately conclude that the correction factor always satisfies $\hat{\alpha}=\frac{\sigma_{z}^{2}}{h_{n}}\hat{s}<1$, that is, ignoring the interference, our reliability metric is too ``optimistic'' and must be scaled down. On the other hand, since $\hat{s}> \hat{s}^{0}$, if the mismatched L-value is calculated using  the Gaussian approximation of the interference, that is using \eqref{Lcorr.0}, the correction would be $\hat{\alpha}>1$. That is, the Gaussian approximation is too ``pessimistic''.

We show in Fig.~\ref{Fig:Corr.Fact} the values of the optimal correction factors for different values of SNR and SIR. For high SNR, as predicted by \eqref{alpha.inf}, $\hat{\alpha}=\hat{\alpha}^{\infty}=1-g_{n}=1-10^{-SIR/20}$; for example, when $SIR=6$~dB, $\hat{\alpha}=0.5$. We also show the value of $\hat{\alpha}^\tr{GMI}$, cf.~\eqref{GMI}  and we see that it is quite close to  $\hat{\alpha}$. Both factors are increasing with SINR as the case $SINR\rightarrow\infty$ corresponds to the assumed absence of the interference, that is, $\hat{\alpha}\rightarrow 1$.

It is also interesting to compare the obtained correction factors to those that might be obtained minimizing the actual PEP. To make it possible we again analyze the two-state mismatch from Sec.~\ref{Sec:Binary.Mismatch}. That is, we assume that $d_{1}$ L-values affecting PEP are mismatched as in our example, while $d_{2}$ L-values are matched, that is no interference is present during the transmission.

From \eqref{l.tilde.BPSK} we easily deduce the distribution of the mismatched L-values $\tilde{L}$ and the matched ones $L$  as
\begin{align}\label{}
   p_{\tilde{L}| C}(l | 0) &= \frac{1}{2}\left[\Psi\left(l +\tilde{\mu}_{1} ; \sigma^{2} \right)
           + \Psi\left(l + \tilde{\mu}_{2}; \sigma^{2} \right)\right]\nonumber\\
           p_{L| C}(l | 0)&=\Psi(l+\mu_{0};\sigma^{2})
\end{align}
where $\tilde{\mu}_{1}=\frac{2 h\cd (h-g)}{\sigma^{2}_{z}}$, $\tilde{\mu}_{2}=\frac{2 h\cd (h+g)}{\sigma^{2}_{z}}$, $\mu_{0} = \frac{2 h^{2}}{\sigma^{2}_{z}}$,  and $\sigma^{2}=\frac{4 h^{2}}{\sigma^{2}_{z}}$.

Since the convolution of $d_{1}$ corrected mismatched L-values $\tilde{L}^\tr{c}$ with $d_{2}$ matched L-values $L$ is a mixture of Gaussian function, we easily obtain the analytical expression for the PEP of two-state mismatch (2SM)
\begin{align}\label{}
\tr{PEP}^\tr{2SM}(\alpha) = \frac{1}{2^{d}} \sum_{k=0}^{d_{1}} {d_{1} \choose k} Q\left( \frac{(d_{1}-k)\alpha\tilde{\mu}_{1}+k\alpha\tilde{\mu}_{2}+d_{2}\mu_{0}}{ \sigma\sqrt{d_{1}\alpha^{2}+d_{2}}} \right),
\end{align}
and we define 
\begin{align}\label{alpha.PEP}
\hat{\alpha}^\tr{2SM}=\tr{argmin}_{\alpha}\tr{PEP}^\tr{2SM}(\alpha),
\end{align}
which is shown in Fig.~\ref{Fig:Corr.Fact} for various combinations of $2\leq d_{1}, d_{2}\leq 8$ as, now $\hat{\alpha}^\tr{2SM}$ depends on $d_{1}$ and $d_{2}$. We can appreciate that the optimal values $\hat{\alpha}^\tr{2SM}$ are very close to $\hat{\alpha}$ resulting from  the saddlepoint-based criterion we proposed, which is obtained without restrictive assumptions on the two-state structure of the mismatch.

For completeness we also show the results of WLSF defined in \eqref{WLSF}. We can expect it to provide reasonable results when $f^\tr{c}(l)$ is almost linear, that is when the pdf $p_{\tilde{L}_{n}|C_{n}}(l|0)$ is close to Gaussian. This happens when interference is dominated by the noise, i.e, for low SNR and high SIR and then, as we can see in Fig.~\ref{Fig:Corr.Fact},  $\hat{\alpha}^\tr{WLSF}$ is close to the  PEP-minimizing values $\alpha^\tr{2SM}$. On the other hand, when SNR is high and SIR low, they results obtained are  far from the optimal values. 

Now, to take our solution out of the PEP-related consideration, and to verify how the correction affects the performance of a practical decoder, we consider a case where a block of $N_\tr{b}=1000$ bits is encoded using the convolutional encoder $\set{15,17}_{8}$ of rate $\frac{1}{2}$  \cite{Frenger99} and the turbo code $\set{1,15/13}_{8}$ \cite{Vogt00} with rate $\frac{3}{4}$ (obtained via puncturing of the parity bits). 

\begin{figure}[t]
\psfrag{xlabel}[c][c]{$\ov{SNR}$}
\psfrag{ylabel}[c][c]{BER}
\psfrag{XXXXXXX1}{$\alpha=1$}
\psfrag{XXXXX2}{$\hat{\alpha}$}
\psfrag{XXXXX3}{$\hat{\alpha}^\tr{GMI}$}
\psfrag{XXXXX4}{true L-values}
\psfrag{XXXXX5}{$\hat{\alpha}^{0}$}
\psfrag{TC}{TC}\psfrag{CC}{CC}
\begin{center}
\scalebox{0.85}{\includegraphics[width=1\linewidth]{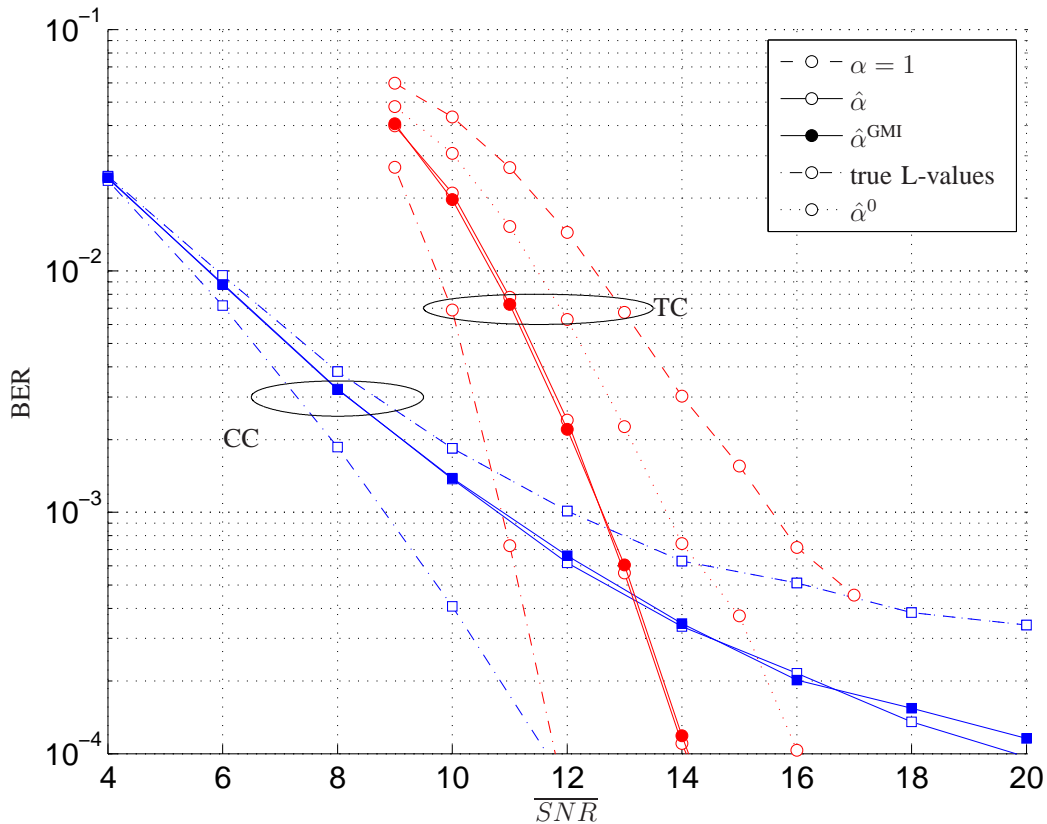}}

\caption{Simulation results for the rate-$\frac{1}{2}$ convolution code (CC) and rate-$\frac{3}{4}$ turbo code (TC) obtained using the L-values without correction \eqref{l.tilde.BPSK} (i.e., $\alpha=1$), the L-values with optimal correction we propose (calculated solving \eqref{SP.eq.r}), the true L-values \eqref{L.value-sumexp}. The results obtained from L-values corrected  using $\hat{\alpha}^\tr{GMI}$ are shown with the filled markers. For CC we use ML decoding, so the results based on the Gaussian model of the interferences ($\hat{\alpha}^{0}$), \eqref{Lcorr.0} are identical to those obtained without the correction as discussion at the end of Sec.~\ref{Sec:Num.Results}.}\label{Fig:BER.BPSK}
\end{center}
\end{figure}

We recall that for identically distributed L-values, the performance of the ML decoder cannot be improved via linear correction. Thus, to show the eventual advantage of the correction, we assume that the channel gains $h_{n}$ are unitary-energy, Rayleigh variables, so the \emph{average} SNR $\ov{SNR}=1/N_{0}$ is used to characterize the channel. The correction factor has to be found for each value of $h_{n}$ that is assumed perfectly known at the receiver. The average signal-to-interference is set to $SIR=6$~dB for CC and $SIR=8$~dB for TC; for these values, the measurable BER results can be presented in the same range of SNR. The results of decoding (Viterbi decoding for the convolutional code and turbo decoding with five iterations for turbo-code) in terms of bit-error rate (BER) are shown in Fig.~\ref{Fig:BER.BPSK}. We also show the results of the decoding using true L-values, i.e., L-values obtained via \eqref{L.value-sumexp}. 

The comparison with the GMI-based correction, is in order even if, as shown Fig.~\ref{Fig:Corr.Fact}, the correction factors are similar  to those obtained using our method. Note that, unlike in our method, solving the GMI-based problems \eqref{GMI} or \eqref{cond.GMI.3}, the numerical integration is needed and the solutions $\hat{\alpha}^\tr{GMI}$ turns out to be sensitive to the number of points of the numerical quadrature (we used Gauss-Hermite method with 40-100 points). Due to these numerical issues, beyond $SNR=15$~dB and particularly for large $SIR$ we were not able to find the solution of \eqref{GMI} in the interval $\alpha\in(0,1)$ (where it must belong). These practical aspects also speak in favour of the correction based on the saddlepoint equation \eqref{SP.eq.r}, where no integration was necessary and the solution was readily obtained using \eqref{Newton.iterations}. To go around these implementation problem, for instantaneous SNR $|h_{n}|^{2}/N_{0}$ larger than 15~dB (i.e., where $\hat{\alpha}$ and $\hat{\alpha}^\tr{GMI}$ are quite similar, cf.\ref{Fig:Corr.Fact}) we used  $\hat{\alpha}$ instead of $\hat{\alpha}^\tr{GMI}$. We also opted for an off-line solution: we pre-calculate a table of $\hat{\alpha}^\tr{GMI}$ for many values of SNR and SIR and, during the simulations, for each instantaneous SNR and instantaneous SIR $|h_{n}|^{2}/g_{n}^{2}$ the value of $\hat{\alpha}^\tr{GMI}$ is interpolated from the table.

We can see that the correction results based on our method or on the GMI approach are similar and  bridge partially the gap to the results based on the true L-values. The performance improvement is particularly notable for high average SNR, which  is consistent with the results of Fig.~\ref{Fig:Corr.Fact} where the most significant correction (small values of $\hat{\alpha}$) are obtained for high SNR.

In Fig.~\ref{Fig:BER.BPSK} we also show the results of the correction derived assuming that the interference is Gaussian yielding the correction factor $\hat{\alpha}^{0}=\frac{\sigma_{z}^{2}}{\sigma^{2}_{\tr{N+I},n}}$. Since $\hat{\alpha}^{0}$ is independent of the channel gains $h_{n}$, it is common to all the L-values and thus irrelevant to the performance of ML decoder. For this reason, the results obtained with $\hat{\alpha}^{0}$ and with $\hat{\alpha}$ are identical for CC, where the ML (Viterbi) decoder is used; they are thus not shown in Fig.~\ref{Fig:BER.BPSK}. On the other hand, the turbo decoder, based on the iterative exchange of information between the constituent decoders, depends on the accurate representation of the aposteriori probabilities via the L-values \cite{Nguyen11}. It is, therefore, sensitive to the scaling and the correction with $\hat{\alpha}^{0}$ improves the results comparing to those obtained without correction.

\section{Conclusions}
In this work we propose a new method to find the linear correction of the mismatched L-values. Aiming at the minimization of the probability of errors made by the maximum-likelihood decoder, we find that the correction factor equals to the twice of the saddle point of the cumulant generating function (CGF) of the L-values. Our method is shown to bear similarities to the one based on generalized mutual information proposed before but is simpler to implement because, working in the domain of CGF,  the numerical integration may be avoided.  We illustrate our findings with the analysis of the BPSK transmission in the presence of the interference where the correction factor clearly improves the performance comparing to the mismatched metrics without correction.

\section*{Acknowledgement}
Many thanks to Alex Alvarado, Cambridge University, UK, for a careful and critical reading of the manuscript.

\balance

\bibliography{IEEEabrv,references_all}
\bibliographystyle{IEEEtran}

\end{document}